\documentclass[preprint,12pt,authoryear]{elsarticle}

\usepackage{amssymb}
\usepackage{graphicx}
\usepackage{natbib}
\usepackage{color}
\usepackage[usenames,dvipsnames,svgnames,table]{xcolor}

\newcommand{\ts}{\thinspace}
\newcommand{\GX}{GX{\thinspace}1$+$4}
\newcommand{\ONE}{1E{\thinspace}1740.7$-$2942}
\newcommand{\GRS}{GRS{\thinspace}1758$-$258}

\journal{Nuclear Physics B}

\begin{document}

\begin{frontmatter}



\title{Telescope performance and image simulations of the balloon--borne coded--mask protoMIRAX experiment}

\author{A. V. Penacchioni} \ead{ana.penacchioni@inpe.br} \author{J. Braga}\ead{joao.braga@inpe.br} \author{M. A. Castro} \ead{manuel.castro@inpe.br} \author{F. D'Amico} \ead{flavio.damico@inpe.br}

\address{Instituto Nacional de Pesquisas Espaciais (INPE), Av. dos Astronautas, 1758, CEP 12227-010, S\~ao Jos\'e dos Campos, SP, Brazil. Tel: +55 (12) 3298 7217}



\begin{abstract}
In this work we present the results of imaging simulations performed with the help of the GEANT4 package for the protoMIRAX hard X-ray balloon experiment. The instrumental background was simulated taking into account the various radiation components and their angular dependence, as well as a detailed mass model of the experiment. We modeled the meridian transits of the Crab Nebula and the Galatic Centre region during balloon flights in Brazil ($\sim-23^{\circ}$ of latitude and an altitude of $\sim 40 \ts$ km) and introduced the correspondent spectra as inputs to the imaging simulations. We present images of the Crab and of three sources in the Galactic Centre region: \ONE, \GRS\ and \GX. The results show that the protoMIRAX experiment is capable of making spectral and timing observations of bright hard X-ray sources as well as important imaging demonstrations that will contribute to the design of the MIRAX satellite mission.
\end{abstract}


\end{frontmatter}

\newpage


\section{Introduction}
\label{Introduction}

X and low-energy ($\lesssim$ 1 \ts MeV) $\gamma$-ray photons coming from astrophysical sources interact in the Earth's atmosphere through photoelectric effect or Compton scattering, never reaching the surface. However, at usual stratospheric balloon altitudes ($\sim$40{\ts}km), a major fraction of the X-ray photons with energy greater than $\sim$30{\ts}keV is able to pass through the remaining external atmosphere. This allows the detection and observation of cosmic sources above these energies with instruments carried by balloons. 
In addition to photons coming directly from the astrophysical sources of interest, radiation from the atmosphere and from other cosmic sources (either point-like or diffuse) in the field of view of an X-ray telescope interacts with the instrument and the balloon platform, generating an intense background. Therefore, detailed predictions of the fluxes and spectra of cosmic sources are crucially dependent upon an accurate knowledge of this background. At balloon altitudes, this is a non-trivial task that has to take into account the several particle and photon fields incident upon the experiment and all the interactions that occur in the detectors and surrounding materials, as well as a detailed mass model of the instrument. By knowing the particle kinds, both primary and secondary, their energy spectra, and by following the secondary emission they produce when they interact, we can simulate the energy depositions on the detectors and identify the contribution of each background component. This is essential to estimate the instrument sensitivity as a function of energy.

In this work, we first show the results of background calculations necessary to make realistic simulations of the astrophysical observations to be carried out by the protoMIRAX experiment. protoMIRAX consists of a wide field coded-mask hard X-ray imager that will operate in a stabilized balloon gondola with fine pointing capability. Besides being a prototype designed to test several subsystems of the MIRAX satellite experiment \citep{2004AdSpR..34.2657B,2006AIPC..840....3B} in a near-space environment, protoMIRAX is also capable to carry out spectral observations of bright X-ray sources. 

We then present imaging simulations of known hard X-ray point sources in selected sky fields to be observed by the experiment in the $30$--$200${\ts}keV energy range. We use standard cross-correlation techniques to reconstruct the images with coded masks. Even though coded aperture imaging is inherently a low signal-to-noise-ratio (SNR) technique, since the whole detector area is used to measure both source and background, it remains the most used imaging technique at hard X-ray and low-energy $\gamma$-ray energies due to its relative simplicity of implementation and the capacity to image large fields of view with reasonable angular resolution. Although the NuSTAR mission has extended X-ray optics up to $\sim 80$ keV \citep{2013ApJ...770..103H,2010SPIE.7732E..0TH}, coded masks are still important for hard X-ray and low-energy $\gamma$-ray wide field imaging instruments.

Even though protoMIRAX is chiefly a prototype designed to test several subsystems of the MIRAX satellite experiment in a near-space environment, we show in this work that the experiment is also capable of making spectral observations of selected bright hard X-ray sources and perform important imaging demonstrations that will contribute to the design of the MIRAX mission.

In section \ref{sec:payload} we describe the protoMIRAX experiment, which will be assembled in a gondola to fly on stratospheric balloons. In section \ref{sec:sample} we make a brief description of the nature of the hard X-ray point sources that we have used in our simulations. In section \ref{sec:sim} we describe the procedure we took to simulate the instrumental background using GEANT4. In section \ref{sec:flux} we show the calculations we made to simulate the detected fluxes of the sources during their meridian passages. In section \ref{sec:image} we show the simulated images and discuss their properties. Finally, in section \ref{sec:conclusions} we present the conclusions. 

\section{The protoMIRAX experiment}\label{sec:payload}

The protoMIRAX balloon gondola houses the X-ray camera, which is the imaging unity of the experiment,  and other subsystems. Table \ref{tab:overview} presents an overview of the experiment and its baseline parameters.

\begin{table}[!h]
\centering
\caption{Overview of the ProtoMIRAX balloon mission.}
\label{tab:overview}
\begin{tabular}{cc}
\hspace{\textwidth}\\
\hline
\multicolumn{2}{c}{\textbf{Mission}}\\
\hline
Total mass& $\sim 600$ kg\\
Altitude& $\sim 42$ km at $-23^{\circ}$ latitude\\
Geometrical area & $169$ cm$^2$\\
Effective area @ 50 keV& $50$ cm$^2$ (through mask)\\
\hline
\multicolumn{2}{c}{\textbf{Detector system}}\\
\hline
Single detector dimensions & $10$mm$\times 10$mm$\times 2$mm\\
Material& Cadmium Zinc Telluride (CZT)\\
Number of detectors& 169 ($13 \times 13$)\\
Gap between detectors & $10$ mm\\
Energy range & $30 - 200$ keV\\
Time resolution & 10 $\mu$s\\
\hline
\multicolumn{2}{c}{\textbf{Coded mask}} \\
\hline
Material& Lead\\
Basic pattern& $13 \times 13$ MURA\\
Extended pattern & $2 \times 2$ basic (minus 1 line and 1 column)\\
Element size& $20$mm $\times\ 20$mm $\times\ 1$mm\\
Open fraction& $0.497$\\
Total mask dimensions& $500 \times 500$ mm$^2$\\
Position & $650$ mm from detector plane\\
\hline
\multicolumn{2}{c}{\textbf{Imaging parameters}}\\
\hline
Angular resolution& $1^{\circ}45'$\\
Total (fully-coded) FOV& $21^{\circ} \times 21^{\circ}$ (FWHM = $14.1^{\circ} \times 14.1^{\circ}$)\\
\hline
\end{tabular}
\end{table}

The camera includes a coded mask based on an extended (4$\times$4) pattern of 13{\ts}$\times${\ts}13 Modified Uniformly Redundant Arrays (MURA) \citep{1989ApOpt..28.4344G}, which is placed 650{\ts}mm away from a position-sensitive detector plane. The mask elements are made of lead with 20{\ts}mm{\ts}$\times${\ts}20{\ts}mm in area and 1{\ts}mm thickness. Such a design will allow a $\sim1^{\circ}45'$ geometrical angular resolution and a $\sim 20^{\circ} \times 20^{\circ}$ fully-coded field of view (FOV). 

The detector plane comprises $169$ square (10{\ts}mm{\ts}$\times${\ts}10{\ts}mm) Cadmium Zinc Telluride (CZT) detectors, 2-mm thick, providing a $\sim${\ts}30--200{\ts}keV optimum energy coverage. The lower threshold is defined by atmospheric absorption at balloon altitudes, whereas the upper limit is due to the detector thickness. We have developed an electronics acquisition system for the CZTs and our current figure for energy resolution is $\sim$8{\%}@60{\ts}keV. 

The capabilities of protoMIRAX in the time domain depend upon the strength of the source signals, which in turn depends on the detector area. The electronic time resolution is 10 $\mu$s. For strong sources, we will be able to detect time variability at a level of tens of seconds. According to the GEANT4 simulations, the expected count rate for the detector plane is approximately 50 counts/s. At those rates we will have a very low dead time and no pile-up problems.

\begin{figure}[!h]
\begin{center}
\includegraphics[width=0.7\textwidth]{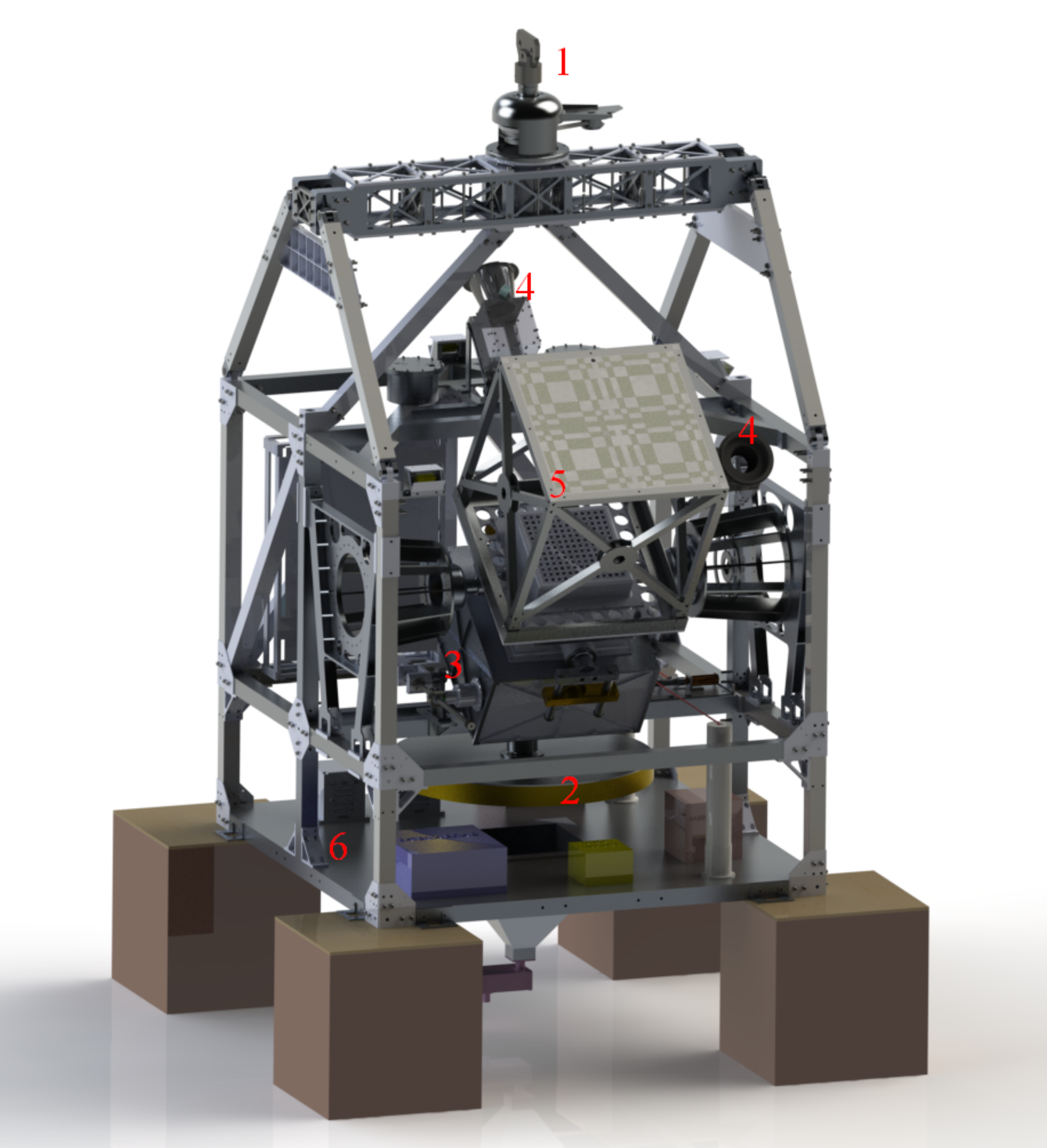}
\caption{Design of the protoMIRAX payload, consisting of: (1) the pivot with a decoupling mechanism for momentum dump; (2) the reaction wheel; (3) the elevation step-motor-controlled actuator; (4) star sensors; (5) the MURA-based coded-mask; and (6) the electronics bay. The gondola is $\sim$2.4{\ts}m high, 1.4{\ts}m $\times$ 1.4{\ts}m in cross section, and its total mass is around 600{\ts}kg}
\end{center}
\label{pmrxgondola}
\end{figure}

The balloon gondola has an attitude control system that provides pointing capability both in azimuth and elevation with an accuracy of a few arc minutes. Fig. {\ts}{\ref{pmrxgondola}} shows a computer design of the protoMIRAX experiment.

\section{The sample of simulated point sources}\label{sec:sample}

\subsection{Crab Nebula}

The well-known Crab Nebula (M1; $\alpha = 5$h $34.5$m; $\delta = +22^{\circ} 01$') is the remnant of a bright Galactic supernova recorded by Chinese astronomers in 1054. It is at a distance of 2.2{\ts} kpc from Earth and today has an apparent magnitude $m_v = 8.6$. The system encompasses a plethora of regions with different structures and emission mechanisms. At the centre of the nebula lies the Crab Pulsar, the only neutron star that emits coherent pulses in phase all the way from radio to $\gamma$-rays, with a period of $P =33${\ts}ms and $\dot{P} = 4.21\times 10^{-13}$. The correspondent spin-down luminosity is $L_{sd} = 4\pi^2 I P^{-3} \dot{P} \sim 5\times 10^{38}${\ts} erg s$^{-1}$, where $I$ is the moment of inertia of the neutron star \citep{2002apa..book.....F}. 
Only a small fraction of $L_ {sd}$ goes into pulsed emission; the majority is carried away by some combination of magnetic dipole radiation and an ultrarelativistic wind. This is the source of the energy that powers the Crab. It is currently believed that a Pulsar Wind Nebula (PWN) efficiently converts the energy of the shocked pulsar wind into the observed synchrotron emission, which accounts for $\sim$26\% of the total $L_{sd}$. For a comprehensive review on the Crab, see \citet{2008ARA&A..46..127H}.

In X-ray astronomy, the Crab has been considered as a standard candle due to its large and nearly constant luminosity. In the literature, the fluxes of X-ray sources in a given energy band are often measured in ``Crabs", meaning the flux of the Crab in the same energy band. In the range of 2 to 10{\ts}keV, 1{\ts}Crab = $2.4\times 10^{-8}$ erg cm$^{-2}$ s$^{-1}$ .

At X-ray energies above $\sim$30{\ts}keV, the Crab is generally the strongest persistent source in the sky and has a diameter of $\sim$1{\ts}arcmin. 

The Crab has experienced some recent flares in the GeV range \citep{2011Sci...331..736T} which were simultaneous or just previous to a particularly active period in hard X-rays \citep{2011ApJ...727L..40W}. However, those surprising variations are of the order a few percent on the 50-100 keV energy range and do not make any significant difference in the estimations made in this work. We will observe the Crab Nebula for flux calibrations and imaging demonstrations.

\subsection{Galactic Centre sources} 

\subsubsection{\ONE}

 \ONE\ is the brightest and hardest persistent X-ray source within a few degrees of the Galactic Centre (GC; $\alpha =17$h $45$m $40.04$s; \,\,\, $\delta=-29^{\circ} 00' 28.1$'') and, due to a similar hard X-ray spectrum and comparable luminosity to Cygnus X-1 \citep{1984SSRv...38..353L}, is classified as a black hole candidate \citep{S.1991}. Because of the two-sided radio jets associated to the source, \ONE\ was dubbed the first ``microquasar" \citep{1992Natur.358..215M,2007ESASP.622..365D}, an X-ray binary whose behavior mimics quasars on a much smaller scale. Since the GC direction has extremely high extinction, with a hydrogen column density around $10^{23}$cm$^{-2}$ \citep{1996ApJ...468..755S}, a counterpart has not yet been identified despite deep searches made in both the optical and IR bands \citep{1991BAAS...23.1392P}. Additional absorption can be attributed to the material of a molecular cloud in which the source is likely embedded \citep{1991Natur.353..234B}. The source was studied recently by our group from soft to hard X-rays up to 200{\ts}keV \citep{CastroAvila:2011:EsCoCo}, showing spectra that can be very well modeled by thermal Comptonization of soft X-ray photons. We will image the Galactic Centre region with protoMIRAX with \ONE\ in the centre of the field-of-view for imaging demonstration and spectral measurements.

\subsubsection{\GRS}

\GRS, also a microquasar, is one of the two brightest X-ray sources near the GC at energies greater than $50$ keV \citep{1991ApJ...383L..49S}. Its hard X-ray spectra and variability are similar to that of Cyg{\ts}X-1 \citep{1996rftu.proc..157K}. \GRS\ also does not have yet an identified counterpart, so its mass and orbital periods are still unknown. The behavior of \GRS\ during the transition between the hard and soft spectral states is markedly different from that of Cyg X-1: there is an observed time delay of $\sim 1${\ts} month between changes in its luminosity and spectral hardness. This suggests that \GRS\ has two separate accretion flows, a thin disk and a halo, which indicates the presence of a low-mass companion powered by Roche lobe overflow \citep{1999ApJ...525..901M}. \GRS\ will be in the field-of-view of our GC observations, so we have included its spectrum here for our imaging simulations with protoMIRAX.

\subsubsection{\GX}

The neutron star low-mass X-ray binary system \GX, the best studied accreting pulsar around the GC, is the prototype of the small but growing subclass of accreting X-ray pulsars called Symbiotic X-ray Binaries (SyXB), by analogy with symbiotic stars, in which a white dwarf accretes from the wind of a M-type giant companion. It is a persistent source, but with strong, irregular flux variations on various timescales and extended high/low states.

The system was discovered in 1970 by a balloon observation at energies above 15 keV showing pulsations with a period of about two minutes \citep{1971ApJ...169L..17L}. Another balloon observation by \citet{1987Ap&SS.137..233J} discovered pulsations at hundreds of keV. Its peculiar spin history has been the object of intensive study \citep{1989PASJ...41....1N}, and \citet{1988Natur.333..746M} have shown a clear transition from spin-up to spin-down behavior at the equilibrium period, characterizing a torque reversal episode. Optical flickering with timescales of minutes were discovered by \citet{1993RMxAA..26Q.113B} in V2116{\ts}Oph, the optical counterpart of \GX, and optical pulsations at the X-ray frequency were discovered by \citet{1997ApJ...482L.171J}. \citet{1999ApJ...526L.105P} have discovered an X-ray frequency modulation of $\sim$304 days in the BATSE/CGRO data, whereas infrared spectroscopy by \citet{2006ApJ...641..479H} have shown a single-line spectroscopic binary orbit with a period of 1161 days,. If this is the orbital period, this system is the widest known low-mass X-ray binary. More recent studies \citep{2012A&A...537A..66G} have used data from {\it Beppo}SAX, INTEGRAL, {\it Fermi\/} and {\it Swift}/BAT to show that the source continues its spin-down trend with a constant change in frequency, and the pulse period has increased by $\sim$50\% over the last three decades.

We will have \GX\ in the field-of-view of the GC protoMIRAX observations. The detection of the source will help to demonstrate the imaging capabilities of the telescope and make an updated measurement of the spin period of this very interesting and peculiar object.

\section{Simulation of the background at balloon altitudes}\label{sec:sim}

To compute the protoMIRAX instrumental background at balloon altitudes we made use of the well-known radiation-interaction code GEANT4, developed by CERN \citep{2006ITNS...53..270A,2003NIMPA.506..250A}. We have built a procedure to take into account the main particles (photons, protons, electrons and neutrons) coming from all directions in the atmospheric environment at balloon altitudes ($\sim40${\ts}km).

In the case of the external electromagnetic radiation, we considered the anisotropic atmospheric X and $\gamma$-ray spectra and the cosmic diffuse contribution coming through the telescope aperture. We have divided the contributions in 4 zenith angle ($z$) ranges from $0^{\circ}$ to $180^{\circ}$, in the $0.024 - 10$ MeV energy range. Table \ref{table:photons} shows the different power-law spectra adopted in each angular range. 

\begin{table}[!h]
\centering
\caption{Incident photon spectra in units of photons cm$^{-2}$ s$^{-1}$ MeV$^{-1}$ in the $0.024 - 10$ MeV energy range, for zenith angles between $0^{\circ}$ and $180^{\circ}$. Taken from \cite{1985NIMPA.239..324G}.}
\label{table:photons}
\begin{tabular}{c|c|c}
\hline
Spectrum                                           & Angular range & Energy range   \\
 & [$^{\circ}$]    & [MeV]          \\
\hline
$2.19 \times 10^2 E^{0.7}$                           & 0 - 65       & $0.024 - 0.035$ \\
$5.16 \times 10^{-2} E^{-1.81}$                       & 0 -65        & $0.035 - 10$    \\
$0.085 \, E^{-1.66}$                                 & 65 -95       & $1 - 10$        \\
$0.140 \, E^{-1.50}$                                 &95 - 130      & $1 - 10$        \\
$0.040 \, E^{-1.45}$                                 &130 - 180     & $1 - 10$        \\
\hline
\end{tabular}
\end{table}

The proton spectrum consists of two components (see Table \ref{table:protons}). The first one (primary) is the most energetic and corresponds to the cosmic rays. They will come only from the upper hemisphere, since they are from cosmic origin. The second component (secondary), less energetic, is the product of the scattering of the primary protons in the atmosphere. Therefore, these protons can come from any direction. We calculated the primary proton distribution in the range $10^4 - 1.5 \times 10^5$ MeV, while for the secondary protons we considered the range $5 - 6600$ MeV. 

\begin{table}[!h]
\centering
\caption{
         Primary and secondary proton spectra as a function of energy. 
         Taken from \citet{2003SSRv..105..285D}
        }
\label{table:protons}
\begin{tabular}{c|c}
\hline
Spectrum                                 & Energy range                          \\
\rm{[protons cm$^{-2}$ s$^{-1}$ MeV$^{-1}$]}& [MeV]                                 \\
\hline
\multicolumn{2}{c}{\textbf{Primary protons}}                                     \\
\hline
$1.3 \times 10^3 \, E^{-2}$               & $10^4$ - $3 \times 10^4$               \\
$1.8 \times 10^6 \, E^{-2.7}$              & $3 \times 10^4$ - $1.5 \times 10^5$    \\
\hline
\multicolumn{2}{c}{\textbf{Secondary protons}}                                    \\
\hline
$1.475 \times 10^{-6} \, E^2$             & 5 - 16                                 \\
$3.78 \times 10^{-4} \, E^0$              & 16- 100                                \\
$6.43 \times 10^{-3} \, E^{-0.61}$         & 100 - 300                               \\
$1.78 \, E^{-1.6}$                        & 300 - 6600                              \\
\hline
\end{tabular}
\end{table}

\begin{table}[!h]
\centering
\caption{Neutron spectra as a function of energy. Taken from \citet{1985NIMPA.239..324G}}
\label{table:neutrons}
\begin{tabular}{c|c}
\hline
Spectrum                                  &Energy range  \\
\rm{[neutrons cm$^{-2}$ s$^{-1}$ MeV$^{-1}]$}& [MeV]        \\
\hline
$0.26 \, E^{-1.3}$                          & 0.8 - 10     \\
$6 \times 10^{-2} \, E^{-0.65}$              & 10 - 100     \\
\hline
\end{tabular}
\end{table}

The electrons appear as a consequence of the interaction of cosmic rays in the atmosphere. Their spectrum in the $1 -10$ MeV energy range is given by:
$$\frac{dN}{dE}\bigg|_{e^-}=1.4 \times 10^{-2} \, E^{-1.8}\, \rm{e^{-} \, cm^{-2}\, sr^{-1}\, s^{-1}\, MeV^{-1}}$$

Table \ref{table:neutrons} shows the neutron spectrum as a function of energy \citep{1985NIMPA.239..324G}. Neutrons are secondary particles, since they appear as a consequence of the interaction of high energy protons and alpha particles in the atmosphere \citep{1973JGR....78.2715A}.

In order to carry out the background simulations using GEANT4, we have built a detailed mass model of the protoMIRAX experiment including the detector plane, the collimator, the coded mask, the support structure and the shielding materials. This model, together with all photon and particle fields described above, were used as inputs to the program.

GEANT4 takes into account all the physical processes by which particles interact with matter. The user can choose which processes are going to be considered in a given simulation. Those processes include pair production, photoelectric effect, Compton scattering and Rayleigh scattering for photons, multiple scattering, ionization, Bremsstrahlung and Coulomb scattering for electrons, and elastic and inelastic processes and Bertini cascades for protons. The processes considered for neutrons include those for protons, plus neutron capture and fission.

\begin{figure}[!ht]
\centering
\includegraphics[width=1.05\hsize, angle=0]{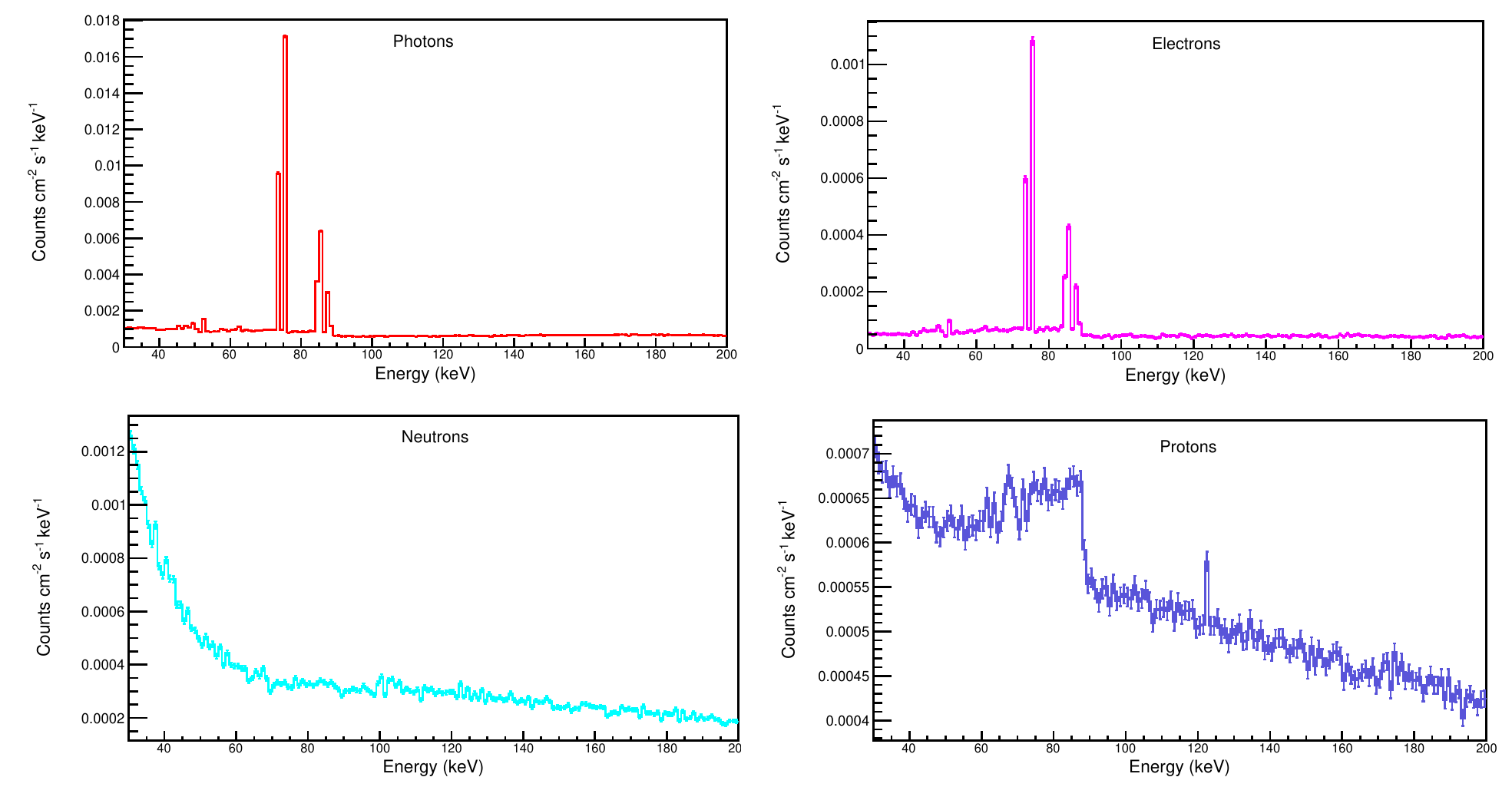}
\caption{Background spectra estimated for each particle species (electrons, photons, protons and neutrons) for protoMIRAX based on GEANT4 simulations at balloon altitudes at a latitude of $-23^{\circ}${\ts}S.  The error bars are 1$\sigma$ statistical uncertainties. We have taken into account the activation of the detector materials in all cases. Six lead fluorescence lines are visible in the photon and electron spectra at 72.8, 74.9, 84.4, 84.9, 87.3 and 88 keV (values taken from http://www.kayelaby.npl.co.uk/atomic\_and\_nuclear\_physics/4\_2/4\_2\_1.html)}.
\label{fig:individual spectrum}
\end{figure}

\begin{figure}[!ht]
\centering
\includegraphics[width=1.0\hsize, angle=0]{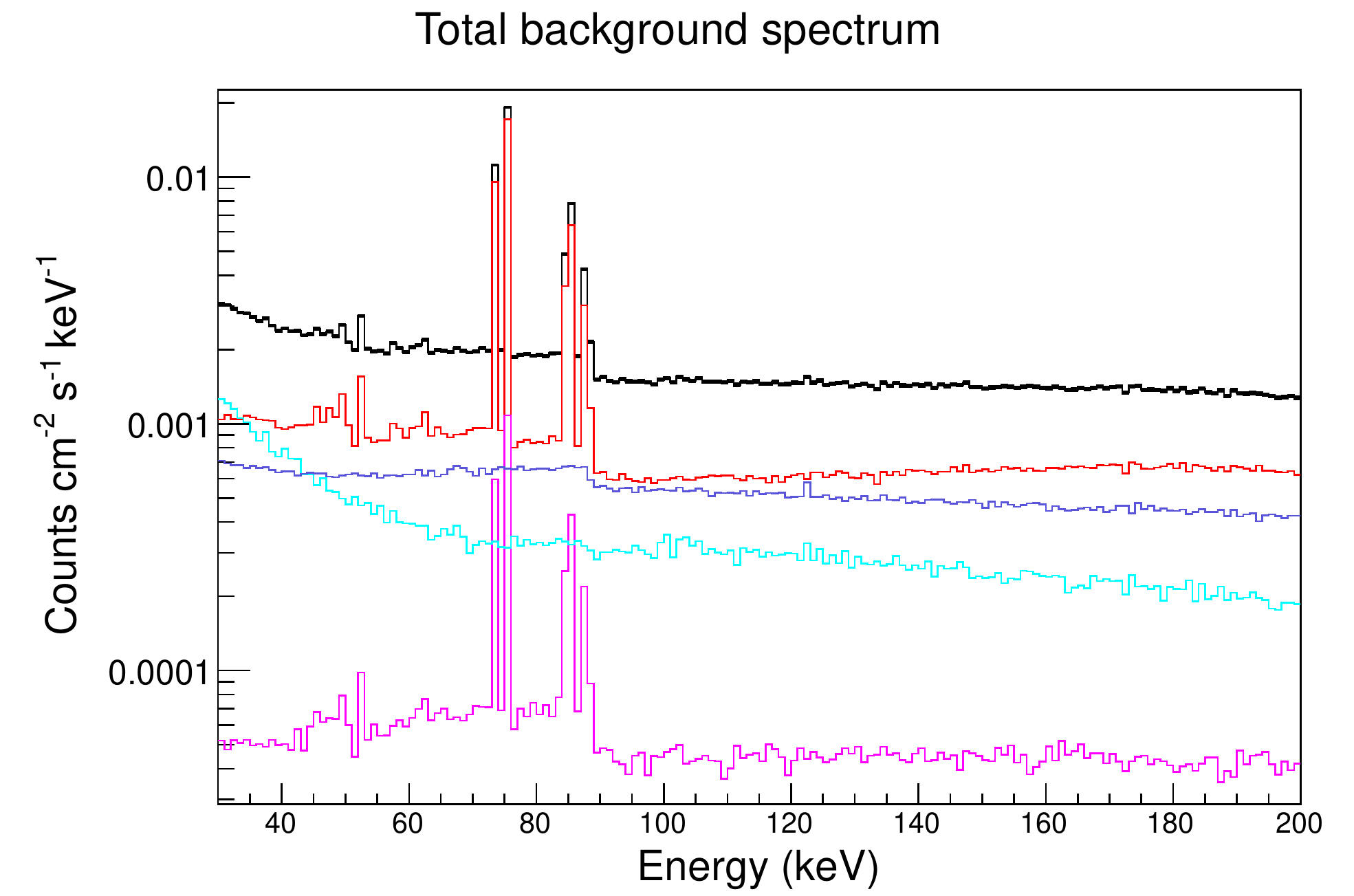}
\caption{Background spectral contributions estimated for protoMIRAX based on GEANT4 simulations at balloon altitudes at a latitude of $-23^{\circ}${\ts}S. Red corresponds to photons, blue to protons, cyan to neutrons, pink to electrons and black represents the sum of all the contributions. The error bars are 1$\sigma$ statistical uncertainties.}
\label{fig:spectrum}
\end{figure}

Fig. \ref{fig:individual spectrum} shows the spectrum of each background particle species considered: photons, electrons, protons and neutrons.
By adding all the interactions that deposit energy onto the protoMIRAX detector plane by direct or indirect hits, considering all the photon and particle fields used in this simulation, we can produce the total background spectrum expected for the protoMIRAX balloon flights. This is shown in Fig. \ref{fig:spectrum} for an integration of $8{\ts}$h at an atmospheric depth of 3 g cm$^{-2}$ at a latitude of $-23^{\circ}$ over Brazil.

\begin{figure}[!ht]
\centering
\includegraphics[width=\hsize, angle=0]{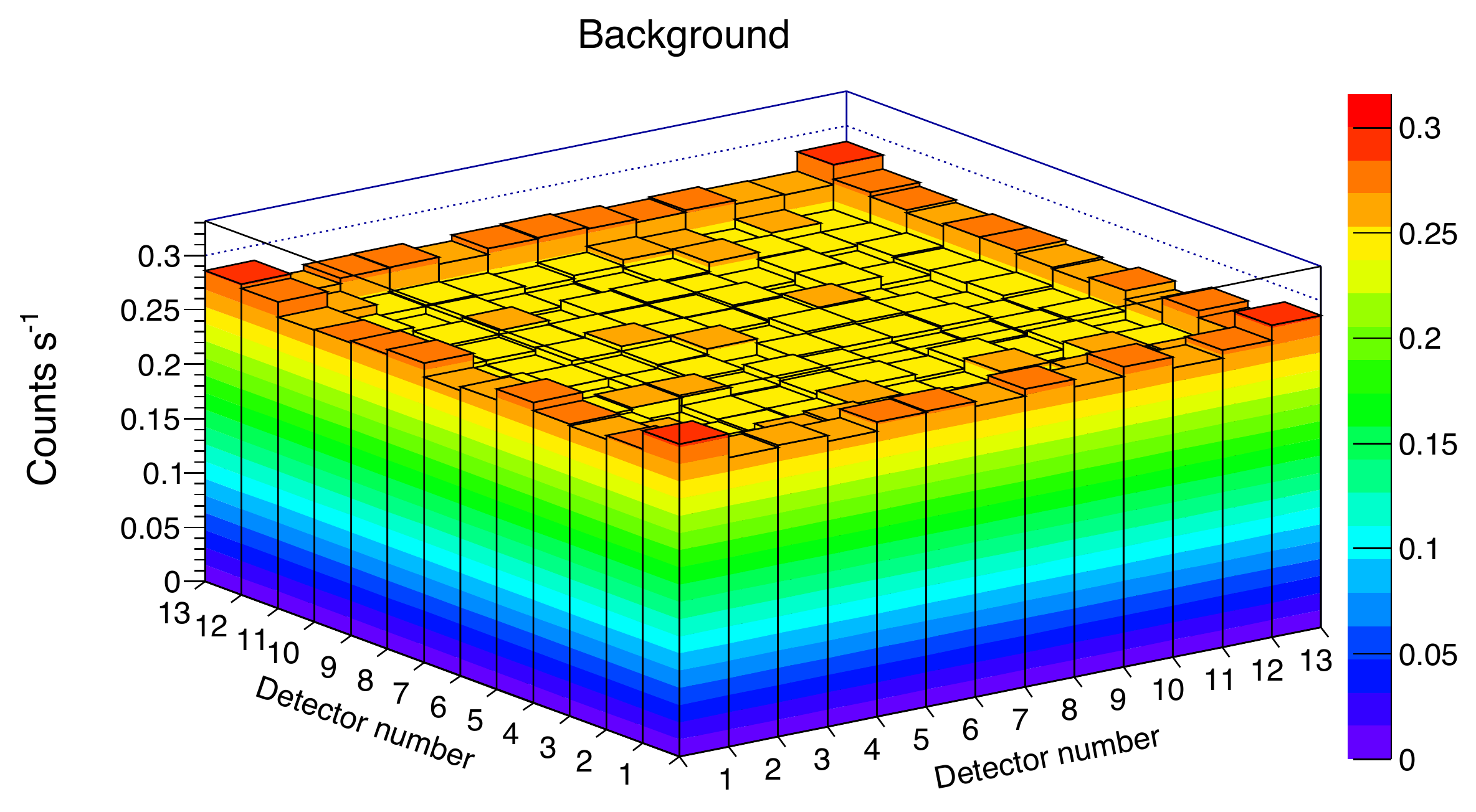} 
\caption{Count distribution of the total background over each pixel on the detection plane in the detector energy range $30 - 200$ keV. The observation time was 8 h at balloon altitudes. Each pixel corresponds to one planar CZT detector (1 cm$^{2}$).}
\label{fig:background total}
\end{figure}

The total number of simulated background counts is 1,453,260, 
which corresponds to $50.46 \pm 0.04$ counts/s. The background drops considerably above 40 keV and stays reasonably flat at a level of $\sim1.6\times 10^{-3}$ counts$^{-2}$ s$^{-1}$ keV$^{-1}$ up to 200 keV. In Fig. \ref{fig:background total} we show the spatial distribution of the background across the detector plane.
The background is reasonably smooth, with slightly higher counting rates near the edges and corners due to geometrical factors. The details of the background simulations, including the relative importance of each input spectrum and the specific properties of the GEANT4 input files we have used ({\it e.g.\/} the mass model), are the subject of another work in preparation that will appear elsewhere soon.

\section{Simulations of the incident fluxes from the point sources}\label{sec:flux}

The X-ray/low-energy $\gamma$-ray photon spectrum of the Crab Nebula is known to be well described by a power-law
\begin{equation}
\centering
F_0=A_0 E^{-\gamma}  \,\,\,\,\,\,\, \rm{photons \, cm^{-2} \,s^{-1} \,\rm{keV}^{-1}},
\end{equation}
with $A_0=14.4$ and $\gamma=2.2$ \citep{2004ESASP.552..815S}. The fluxes at each energy are attenuated due to absorption in the atmosphere, so they vary as
\begin{equation}
\centering
F (E) =F_0 (E) \,e^{-\frac{\mu}{\rho}x \rm{sec}(z)},
\end{equation}
where $\mu/\rho (E)$ is the absorption coefficient of the atmosphere\footnote{http://physics.nist.gov/PhysRefData/XrayMassCoef/ComTab/air.html} at energy $E$, given in cm$^2$ g$^{-1}$, $x=2.7$ \rm{g\,cm$^{-2}$} is the atmospheric depth at the expected balloon altitude ($\sim 40$ km), and $z$ is the zenith angle. 


In order to calculate the total number of counts from the Crab in the protoMIRAX energy range during its meridian passage, we have to integrate over its path in the sky. We considered a 4-hour period divided equally before and after the meridian transit, since the source would be too low in the sky before and after that.

First, we consider the dependence of the zenith angle with time. Let $\phi$ be the geographic latitude ($\phi = -22^{\circ}39'$ at Cachoeira Paulista, S\~ao Paulo, Brazil), $\delta = +\,22^{\circ}01'$ the declination of the source, $\alpha = 05\rm{h} \,34.5\rm{m}$ the right ascension and $H=t-\alpha$ the hour-angle, $t$ being the sidereal time (LST). The zenith angle $z$ is given by
\begin{equation}
\cos{z(t)}=\sin{\phi}\sin{\delta}+\cos{\phi}\cos{\delta}\cos{{(t-\alpha)}}.
\end{equation}
Given the expression for $z(t)$, we can compute the number of photons $N$ coming from the Crab reaching the detector of area $S$ in a time interval $T$ in the detector energy range $E_{min}-E_{max}$. It will be
\begin{equation}
\label{N}
N=\int_0^T \int_{E_{\rm min}}^{E_{\rm max}} \int_S A_0 E^{-\gamma} \rm{exp}\left[{-\frac{\mu}{\rho}(E) \,x\,  \rm{sec} \,z}\right] dE\, dt \,dS.
\end{equation}

For the protoMIRAX experiment, the total energy range is 30 to 200{\ts}keV and the area to be considered for this calculation is $\sim 84$ cm$^2$, since this corresponds to the number of detectors that are not covered by the closed elements of the mask during an observation of a point source. This is the case because, for the 13$\times$13-element MURA pattern used (see section 5), 85 cells are closed and 84 are open. For the Crab simulation, we considered it to be at the centre of the field-of-view, so there are no losses due to the collimator response.

We performed the integration numerically by using sufficiently small energy and time bins so as to get a good approximation, given the variations of the spectrum with energy and of $z$ with time. After calculating the number of counts $N$ for each energy bin, we can run a GEANT4 simulation by shining the corresponding absorbed Crab spectrum over the protoMIRAX mass model. 

\begin{figure}[!h]
\centering
\includegraphics[width=\hsize,angle=0]{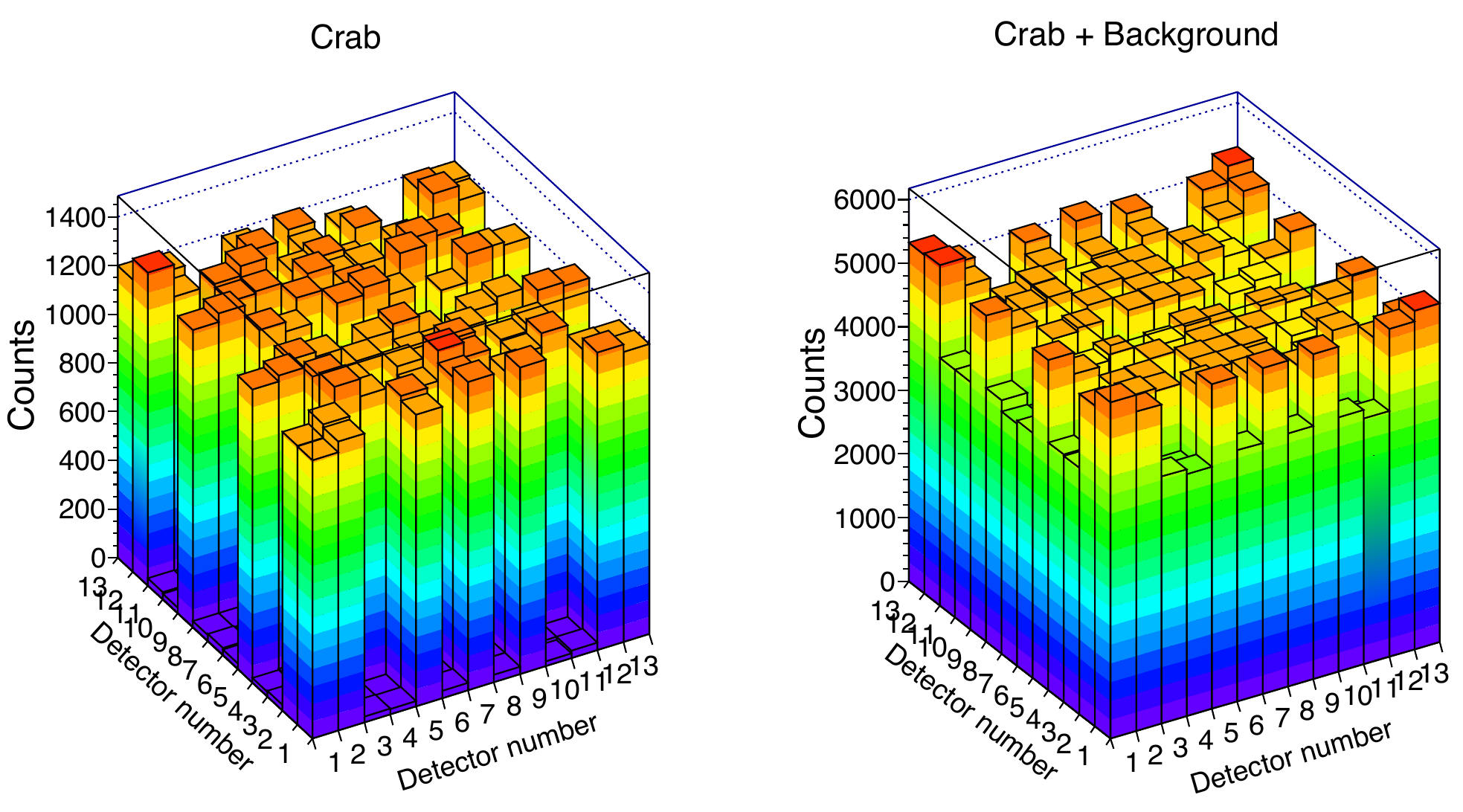}
\caption{
         Crab (left) and total (Crab $+$ background) (right) count distribution over the detection 
         plane in the detector energy range $30 - 200$ keV. The observation time was 4 h.}
\label{fig:diagrama de sombras crab y tot}
\end{figure}

Fig. \ref{fig:diagrama de sombras crab y tot} (left) shows a map of the total counts coming from the Crab in each detector for the 4-hour meridian passage simulation. This Crab map is actually a ``shadow" of the mask projected onto the detector plane, since the flux is spatially modulated by the mask (see section 5). Each pixel of the map corresponds to one detector. The total number of counts due to the Crab is 103331. 
When we add a 4-hour simulated background, we get the complete shadowgram shown in Fig.\ref{fig:diagrama de sombras crab y tot} (right). 

Following the same procedure as the one used for the Crab Nebula, we considered the emission from three sources near the GC: \ONE, \GX\ and \GRS, as described in section \ref{sec:sample}. The parameters of the photon spectra of each of these sources are presented in Table \ref{tabla:fuentesGC}. We have used these spectra and the zenith angle variation of the GC direction to perform the integral of equation (\ref{N}) for each source.

\begin{table}[!ht]
\centering
\caption{Photon spectra of the source sample near the GC. The spectra are in units of photons cm$^{-2}$ s$^{-1}$ keV$^{-1}$}
\label{tabla:fuentesGC}
\begin{tabular}{ccc}
\hline
Source         & Spectrum                                   & Reference                         \\
\hline
\ONE & $10^{-4} \, \frac{E}{100 \rm{keV}}^{-1.35}$           & \citet{1995Grebenev} \\
\GRS & $4.6 \times 10^{-5} \, \frac{E}{100 \rm{keV}}^{-1.8}$ & Sunyaev.et al. (1991b)\\
\GX & $5.1 \times 10^{-4} \, \frac{E}{30 \rm{keV}}^{-1.9}$  & \citet{1991AdSpR..11...35D} \\
\hline
\end{tabular}
\end{table}

\begin{figure}[!h]
\centering
\includegraphics[width=\hsize,angle=0]{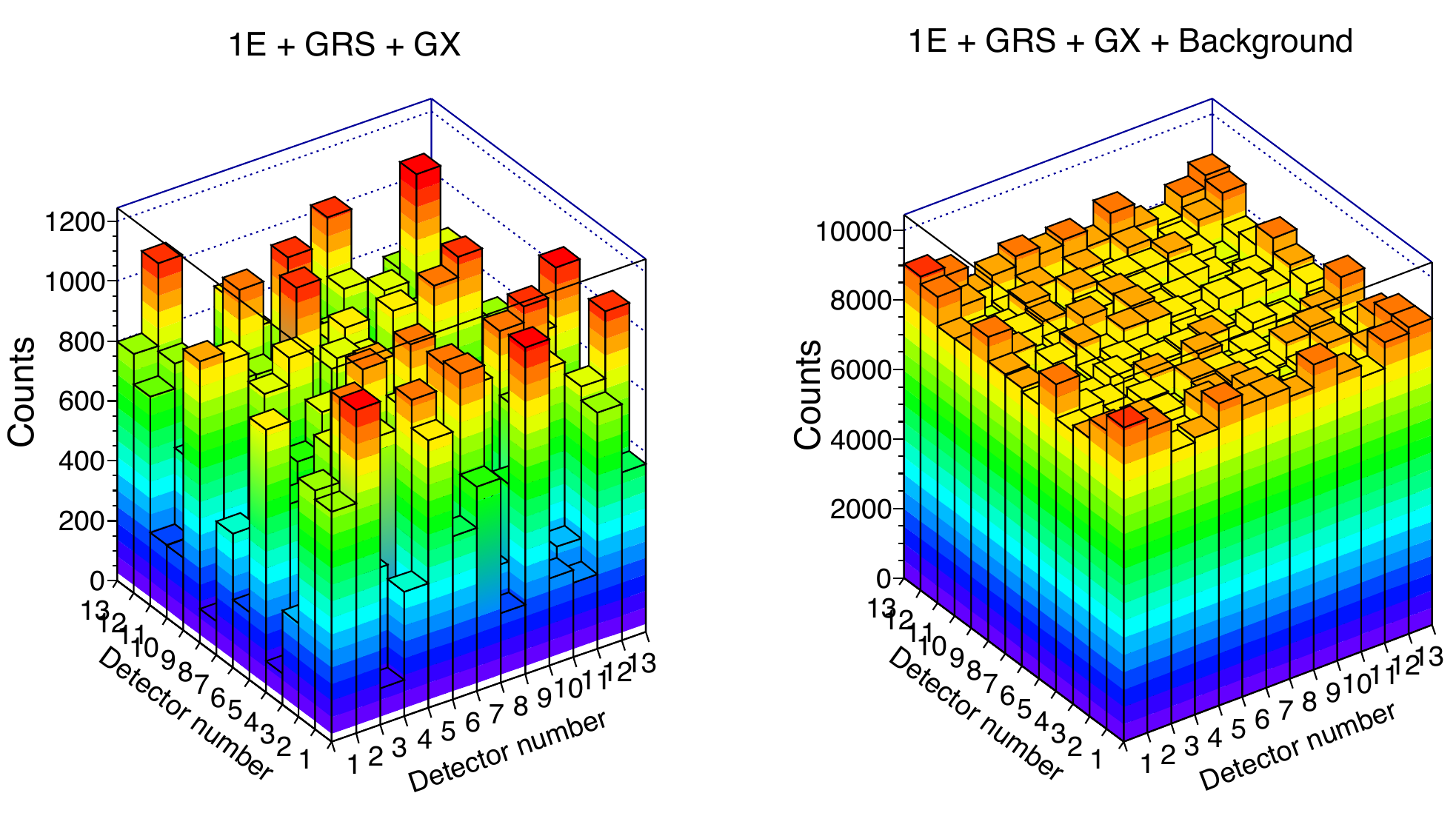}
\caption{The three sources we studied in the GC (left) and the total counts distribution (3 sources $+$ background) (right) over the detection plane in the detector energy range $30 - 200$ keV. The observation time was 8 h.}
\label{fig:diagrama de sombras 3 fuentes}
\end{figure}

In this case, since the GC transits will occur at $\lesssim 6^{\circ}$ from the zenith at the latitudes of the protoMIRAX programmed balloon flights, we have considered integration times of 4 hours before and after the transits. In order to maximize the signal-to-noise ratio (SNR) of \ONE, we have centralized in the image in this source. Fig. \ref{fig:diagrama de sombras 3 fuentes} shows the correspondent shadowgrams for the GC simulation. It is interesting to note that the shadowgram, in this case, bears almost no resemblance to the mask pattern (see section 5), since the overlapping shadowgrams from the three point sources interfere in a complex way.

It is noteworthy that in the actual observations we will have to take into account the sky rotation in the field-of-view during the integration times, due to Earth's rotation. For this simulations, we assumed that this effect will be properly dealt with by rotating the shadowgrams obtained during short integration times, during which the source positions will change by a small angle compared to the instrument resolution, and the rotated shadowgrams will be stacked and added up with the proper orientations.

\section{Image reconstruction}\label{sec:image}

In a coded-mask experiment, the mask is composed of elements which are transparent (open elements) or opaque (closed elements) to radiation. Photons hitting an open element will pass through the mask and reach a position-sensitive detector plane, and photons hitting a closed element will be absorbed. In this way, a pattern of mask ``shadows'' is projected on the detector plane, which actually corresponds to the superposition of many images of each object (one image for each open element). In order to determine the intensity and position of each observed object in the field of view, a decoding and reprocessing of the shadowgram is needed through the implementation of numerical algorithms \citep{1987SSRv...45..349C}. This can be a very complex process in the general case of extended sources at different distances from the instrument. In the present case of point sources at infinity, the reconstruction procedure is somewhat simplified. Essentially, the reconstructed image is the cross-correlation between the detector count distribution map and a decoding function that mimics the mask pattern \citep{1978ApOpt..17..337F}.

\begin{figure}[!ht]
\centering
\includegraphics[width=0.6\hsize,angle=0]{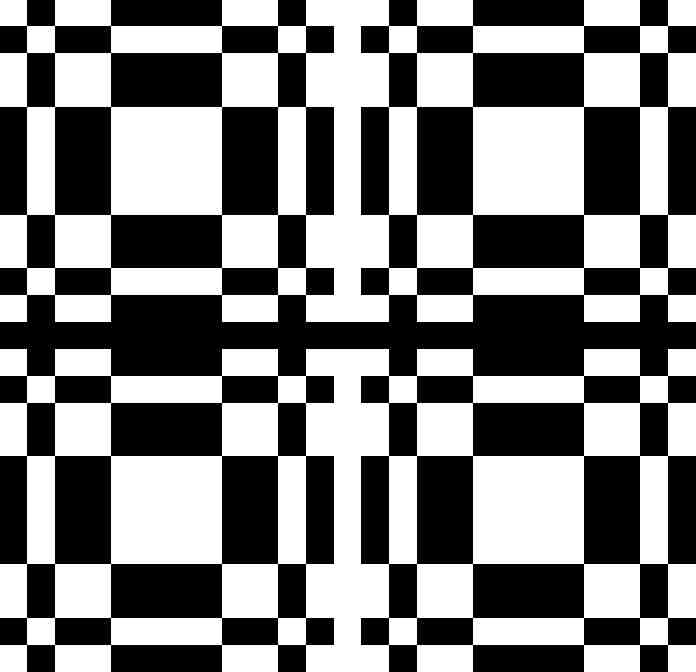}
\caption{The coded mask pattern used by protoMIRAX. It is a cyclic ($4\times4$) repetition of a $13\times13$ MURA pattern, minus one line and one column. The basic cells are made of 1mm-thick lead and measure 10 mm $\times$ 10mm.}
\label{fig:coded mask}
\end{figure}

The coded mask used by protoMIRAX is shown in Fig. \ref{fig:coded mask}. It is a cyclic ($4\times4$) repetition of a $13 \times13$ MURA pattern \citep{1989ApOpt..28.4344G}, minus one line and one column. By using a MURA pattern we get ``perfect'' imaging (no intrinsic noise) and the incomplete repetition of the basic pattern in all directions ensures that we have no ambiguities in the locations of the sources in the field-of-view, since each object always projects a distinct shadow on the detector plane. Since the collimators completely block X rays coming from directions not covered by the mask, we have no partially coded field-of-view in this experiment.

Based on Poisson statistics, the signal-to-noise ratio (SNR) of a peak $(i,j)$ in a reconstructed coded-mask image, which determines the sensitivity of these systems, can be shown \citep{1989ApOpt..28.4344G} to be given by

\begin{equation}
SNR_{ij} = \frac{N_s}{\sqrt{N_s + N_T}} \, , 
\end{equation}
where $N_s$ is the number of net source counts (corresponding to source photons that came through the mask openings and hit the detectors), and $N_T$ is the total number of detector counts, both for the same integration time.

\begin{figure}[!ht]
\centering
\includegraphics[width=\hsize, angle=0]{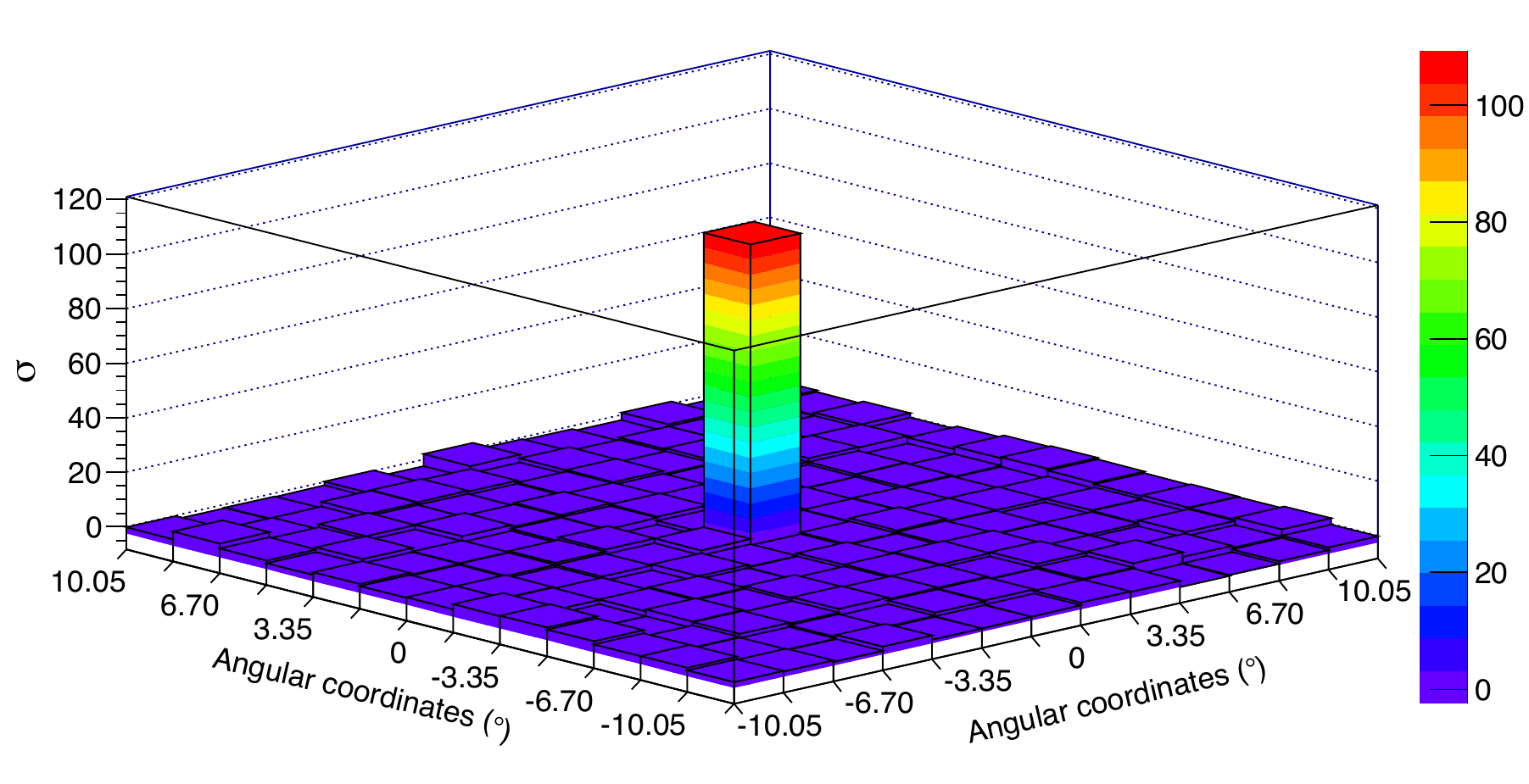} 
\caption{Simulated image of the Crab region as seen by protoMIRAX for 4 hours at balloon altitudes. The Crab is detected at a 109${\sigma}$ level.}
\label{fig:reconstruccion}
\end{figure}

We show in Fig. \ref{fig:reconstruccion} the reconstruction of a simulated image of the Crab Nebula region produced by cross-correlating the shadowgram of Fig.\ref{fig:diagrama de sombras crab y tot} (right) with the mask pattern. The total field-of-view of the image is $21.7^{\circ}$ by $21.7^{\circ}$ and the Crab is detected in one sky-bin of $1^{\circ}45'$. By dividing the peak value by the standard deviation of the off-peak region, we get a SNR of 109, which is 91\% of the theoretical value given by equation 5. This result shows the good performance of the imaging system of the protoMIRAX experiment. Also, our results indicate that we will be able to make a 5$\sigma$ detection of $\sim$1-Crab sources in 30{\ts}s, which will be important to study high-resolution time variability of transient sources.

For the Galactic Centre region simulation, we considered \ONE\ to be in the centre of the field-of-view. The angular distance from this source to \GX\ is $\theta = 5.65^{\circ}$, and to \GRS\ is $\theta = 5.53^{\circ}$. These separations are significantly greater than the protoMIRAX angular resolution, which shows that the instrument is capable of measuring hard X-ray spectra and light curves in relatively crowded fields like the GC.

\begin{figure}[!ht]
\centering
\includegraphics[width=\hsize, angle=0]{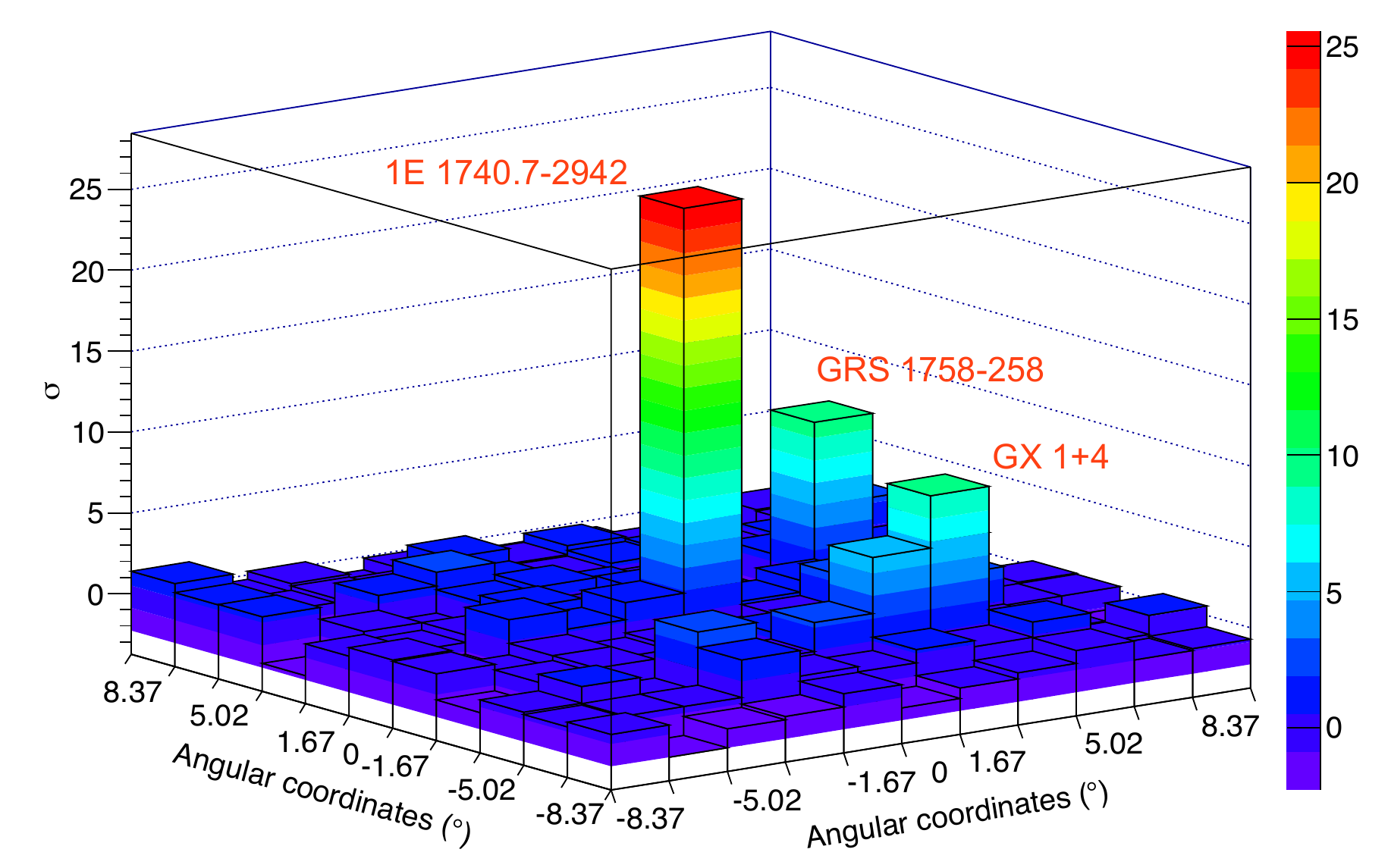} 
\caption{Simulated image of the Galactic Centre region as seen by protoMIRAX for 8 hours. The SNRs for \ONE, \GRS\  and \GX\ are, respectively, 26, 10 and 9. We can see that \GX\ counts are distributed in two sky bins.}
\label{fig:reconstruccion3}
\end{figure}

Fig. \ref{fig:reconstruccion3} shows the reconstructed image of the GC region. To calculate the SNR of each pixel, we first made the identification of the peaks by selecting pixels above 3 standard deviations of the entire image, after subtracting the DC level. We then calculated the SNRs by dividing the peak values by the standard deviation of the off-peaks region. The detection levels are approximately 60\% of the purely statistical values. The signal loss is a combination of 3 effects: non-uniformity of the background across the detector plane (see Fig. 3), the collimator response and the existence of gaps between detectors.

\section{Conclusions}\label{sec:conclusions}

In this work we presented results of simulated images of regions with bright hard X-ray sources with the protoMIRAX balloon experiment. protoMIRAX is a prototype designed to be a scientific and technological pathfinder for the satellite wide-field hard X-ray imager MIRAX.

We introduced the results of GEANT4 simulations of the physical conditions into which the proto-MIRAX experiment will be subject to in its stratospheric balloon flights, in order to predict the performance of the telescope during the observation of bright cosmic sources in hard X-rays. We modeled the instrumental background expected during flights at altitudes of $\sim 40$ km over Southeast Brazil ($\sim -23^{\circ}$ latitude) by simulating the interaction of incident radiation and particles (photons, electrons, protons and neutrons), cosmic and atmospheric, with the experiment materials. 

We then produced images of two regions which will be observed by protoMIRAX in its first balloon flights for calibrations and imaging demonstrations: the Crab Nebula and a Galactic Centre field with 3 bright sources: \ONE, \GRS\ and \GX. The expected fluxes of the sources at balloon altitudes, in several energy bands, were added to the background in order to produce shadowgrams of event distributions over the detector plane.

We simulated the Crab Nebula image considering a meridian passage of 4 hours, during which the highest source elevation will be 45$^{\circ}$. We have shown that we will be able to reach a very good SNR of 109, which will make the Crab a reliable calibration source for the experiment. 

In the case of the GC, the three sources lie near the centre of the field of view and were imaged together, with \ONE\ in the centre. The results show that we can make observations of crowded fields like the GC with protoMIRAX and provide flux and spectral information without source confusion. 

The protoMIRAX mission, as other scientific balloon missions, plays a very important role in testing new detector technology and imaging systems in a near-space environment. As a pathfinder experiment for the MIRAX satellite mission, protoMIRAX is an invaluable source of information for technology development and demonstration.

%
\section{Acknowledgments}
We thank Luiz Reitano, Fernando G. Blanco and S\'ergio Amir\'agile for invaluable technical support. A.V.P. acknowledges the support by the international Cooperation Program CAPES-ICRANET financed by CAPES - Brazilian Federal Agency for Support and Evaluation of Graduate Education within the Ministry of Education of Brazil. We thank FINEP for financial support under Conv\^enio 01.10.0233.00. We also thank CNPq and FAPESP, Brazil, for support under INCT Estudo do Espa\c{c}o.



 \bibliographystyle{elsarticle-harv} 
 \bibliography{references.bib}





\end{document}